\documentclass[
sor,
jor,
amsmath,amssymb,
preprint,
author-numerical,
]{revtex4-1}
\usepackage{natbib}
\usepackage{graphicx}
\usepackage{dcolumn}
\usepackage{bm}
\usepackage{color}
\usepackage{url}
\usepackage{booktabs}
\usepackage{multirow}
\usepackage{amsmath,amssymb}
\usepackage[bookmarks=false]{hyperref}
\hypersetup{colorlinks,allcolors=black}

\begin{document}

\preprint{}

\title[]{Modeling the spread of fake news on Twitter}

\author{Taichi Murayama}
 \email{murayama.taichi.mk1@is.naist.jp}
 \affiliation{Nara Institute of Science and Technology (NAIST)}
 
\author{Shoko Wakamiya}%
 \email{wakamiya@is.naist.jp}
 \affiliation{Nara Institute of Science and Technology (NAIST)}

\author{Eiji Aramaki}%
 \email{aramaki@is.naist.jp}
 \affiliation{Nara Institute of Science and Technology (NAIST)}
 
\author{Ryota Kobayashi}%
 \email{r-koba@k.u-tokyo.ac.jp}
 \affiliation{The University of Tokyo}
 \affiliation{JST PRESTO}

\begin{abstract}
Fake news can have a significant negative impact on society 
because of the growing use of mobile devices and the worldwide increase in Internet access. 
It is therefore essential to develop a simple mathematical model to understand the online dissemination of fake news. 
In this study, we propose a point process model of the spread of fake news on Twitter. 
The proposed model describes the spread of a fake news item as a two-stage process: initially, fake news spreads 
as a piece of ordinary news; 
then, when most users start recognizing the falsity of the news item, 
that itself spreads as another news story. 
We validate this model using two datasets of fake news items spread on Twitter. 
We show that the proposed model is superior to the current state-of-the-art methods in accurately predicting the evolution of the spread of a fake news item. 
Moreover, a text analysis suggests that our model appropriately infers the correction time, i.e., the moment when Twitter users start realizing the falsity of the news item. 
The proposed model contributes to understanding the dynamics of the spread of fake news on social media. 
Its ability to extract a compact representation of the spreading pattern could be useful in the detection and mitigation of fake news. 
\end{abstract}


\maketitle

\section{Introduction}
As smartphones become widespread, 
people are increasingly seeking and consuming news from social media rather than from the traditional media (e.g., newspapers and TV). 
Social media has enabled us to share various types of information and to discuss it with other readers. 
However, it also seems to have become a hotbed of fake news with potentially negative influences on society. 
For example, Carvalho et al.~\cite{united_fake} found that a false report of United Airlines parent company's bankruptcy in 2008 caused the company's stock price to drop by 76\% in a few minutes; it closed at 11\% below the previous day's close, with a negative effect persisting for more than six days. 
In the field of politics, Bovet and Makse~\cite{2016election2} found that 25\% of the news outlets linked from tweets before the 2016 U.S. presidential election were either fake or extremely biased, and their causal analysis suggests that the activities of Trump's supporters influenced the activities of the top fake news spreaders. 
In addition to stock markets and elections, fake news has emerged for other events, including natural disasters such as the East Japan Great Earthquake in 2011 ~\cite{tohoku1,hashimoto2020}, 
often facilitating widespread panic or criminal activities ~\cite{pizzagate}.

In this study, we investigate the question of how fake news spreads on Twitter. 
This question is relevant to an important research question in social science: how does unreliable information or a rumor diffuses in society? 
It also has practical implications for fake news detection and mitigation~\cite{fake_survey1,fake_survey2}. 
Previous studies mainly focused on the path taken by fake news items as they spread on social networks~\cite{Vosoughi2018,Zhao2020}, 
which clarified the structural aspects of the spread. 
However, little is known about the temporal or dynamic aspects of how fake news spreads online.

Here we focus on Twitter and assume that fake news spreads as a two-stage process. 
In the first stage, a fake news item spreads as an ordinary news story. 
The second stage occurs after a correction time when most users realize the falsity of the news story. 
Then, the information regarding that falsehood spreads as another news story. 
We formulate this assumption by extending the Time-Dependent Hawkes process (TiDeH)~\cite{tideh}, a state-of-the-art model for predicting re-sharing dynamics on Twitter. 
To validate the proposed model, we compiled two datasets of fake news items from Twitter.   

The contribution of this study is summarized as follows: 
\begin{itemize}
    \item   We propose a simple point process model based on the assumption that fake news spreads as a two-stage process. 
    \item   We evaluate the predictive performance of the proposed model, which demonstrates the effectiveness of the model.
    \item   We conduct a text mining analysis to validate the assumption of the proposed model. 
\end{itemize}

\section{Related Work} 
Predicting future popularity of online content has been studied extensively~\cite{twitter_predict_analysis,popular_survey}. 
A standard approach for predicting popularity is to apply a machine learning framework, such that the prediction problem can be formulated as a classification~\cite{cascade_predict,rt_to_win} or regression~\cite{predict_online} task. 
Another approach to the prediction problem is to develop a temporal model and fit the model parameters using a training dataset. This approach consists of two types of models: time series and point process models. 
A time series model describes the number of posts in a fixed window. For example, Matsubara et al.~\cite{spikem} proposed SpikeM to  reproduce temporal activities on blogs, Google Trends, and Twitter. In addition, Proskurnia et al.~\cite{predict_sequence} proposed a time series model that considers a promotion effect (e.g., promotion through social media and the front page of the petition site) to predict the popularity dynamics of an online petition.  
A point process model describes the posted times in a probabilistic way by incorporating the self-exciting nature of information spreading~\cite{masuda2013,seismic}. 
Point process models have also motivated theoretical studies about the effect of a network structure and event times on the diffusion dynamics~\cite{delvenne2015}. 
Various point process models have been proposed for predicting the final number of re-shares~\cite{seismic,Medvedev2019}
and their temporal pattern~\cite{tideh} on social media. 
Furthermore, these models have been applied to interpret the endogenous and exogenous shocks to the activity on YouTube~\cite{hip} and Twitter~\cite{fujita2018identifying}. 
To the best of our knowledge, the proposed model is the first model incorporating a two-stage process that is an essential characteristic of the spread of fake news. 
Although some studies~\cite{tornberg2018echo} proposed a model 
for the spread of fake news, 
they focused on modeling the qualitative aspects and did not evaluate prediction performances using a real data set.

Our contribution is related to the study of fake news detection. 
There have been numerous attempts to detect fake news and rumors automatically~\cite{fake_survey1,fake_survey2}. 
Typically, fake news is detected based on the textual content. 
For instance, Hassan et al.~\cite{SVM_word} extracted multiple categories of features from the sentences and applied a support vector machine classifier to detect fake news. Rashkin et al.~\cite{lstm_word} developed a long short-term memory (LSTM) neural network model for the fact-checking of news. 
The temporal information of a cascade, e.g., timings of posts and re-shares triggered by a news story, might improve fake news detection performance. 
Kwon et al.~\cite{time2} showed that temporal information improves rumor classification performance. 
It has also been shown that temporal information improves the fake news detection performance~\cite{csi}, rumor stance classification~\cite{hawkes_stance}, source identification of misinformation~\cite{farajtabar2015}, 
and detection of fake retweeting accounts~\cite{hawkeseye}. 
A deep neural network model~\cite{csi} can also incorporate temporal information to improve the fake news detection performance. 
However, a limitation of the neural network model is 
that it can utilize only a part of the temporal information and cannot handle cascades with many user responses. 
The proposed model parameters can be used as a compact representation of temporal information, which helps us overcome this limitation.

\section{Modeling information cascade of fake news post}
\begin{figure}[!h]
\includegraphics[width=12cm]{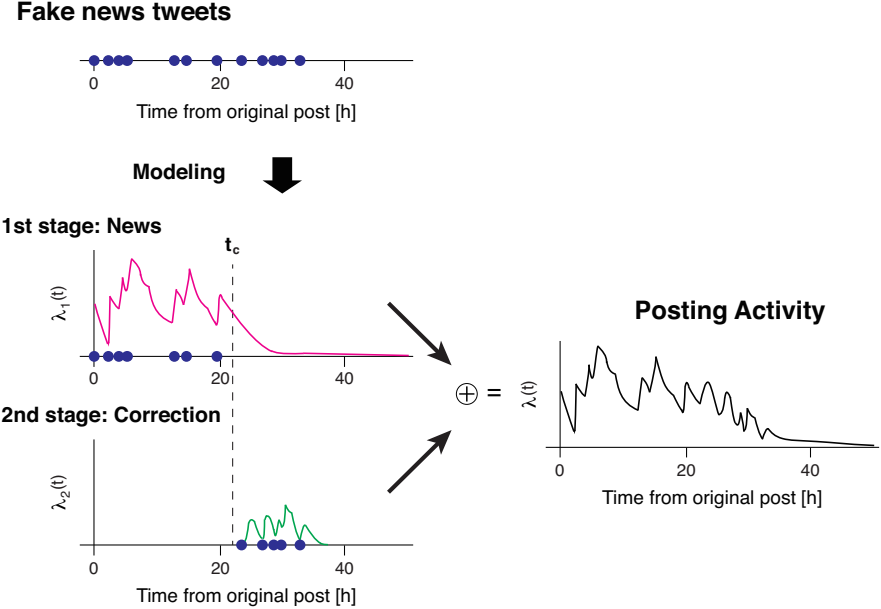}
\caption{Schematic of the proposed model. 
We propose a model that describes how posts or re-shares that are related to a fake news item spread on social media (Fake news tweets). Blue circles represent the time stamp of the tweets.  
The proposed model assumes that the information spread is described as a two-stage process. 
Initially, a fake news item spreads as a novel news story (1st stage). 
After a correction time $t_c$, Twitter users recognize the falsity of the news item. Then, the information that the original news item is false spreads as another news story (2nd stage).
The posting activity related to the fake news $\lambda(t)$ (right: black) is given by the summation of the activity of the two stages (left: magenta and green).
}
\label{overview}
\end{figure}

We develop a point process model for describing the dynamics of the spread of a fake news item. 
A schematic of the proposed model is shown in Fig.~\ref{overview}. 
The proposed model is based on the following two assumptions. 
\begin{itemize}
    \item   Users do not know the falsity of a news item in the early stage. 
    The fake news spreads 
    as an ordinary news story (Fig.~\ref{overview}: 1st stage). 
    \item   Users recognize the falsity of the news item around a correction time $t_c$. The information that the original news is fake spreads as another news story (Fig.~\ref{overview}: 2nd stage). 
\end{itemize} 
In other words, the proposed model assumes that the spread of a fake news item consists of two cascades: 1) the cascade of the original news story and 2) the cascade asserting the falsity of the news story. 
In this study, we use the term {\it cascade} meaning tweets or retweets triggered by a piece of information. 
To describe each cascade, we use the Time-Dependent Hawkes process model, which properly considers the circadian nature of the users and the aging of information.

\vspace{1cm}
\vspace{1cm}

\subsection{Time-Dependent Hawkes process (TiDeH): Model of a single cascade} 
We describe a point process model of a single cascade: the information spreading triggered by a news story.  
In point process models~\cite{daley2003}, the probability of obtaining a post or reshare in a small time interval $[t, t+ \Delta t]$ is written as $\lambda (t) \Delta t$, where $\lambda(t)$ is the instantaneous rate of the cascade, that is, the intensity function. 
The intensity function of the TiDeH model~\cite{tideh} depends on the previous posts in the following manner:   
\begin{equation}
	\lambda_{\rm TiDeH} (t) = p(t) h(t), 
\end{equation}
and the memory function $h(t)$ is defined as follows: 
\begin{equation}
	h(t)= \sum_{i:t_{i} < t} d_{i} \phi(t - t_{i}), 
\end{equation}
where $p(t)$ is the infection rate, $t_i$ is the time of the $i$-th post, and $d_i$ is the number of followers of the $i$-th post. 
The infection rate $p(t)$ incorporates two main properties in the cascade: the circadian rhythm and decay owing to the aging of information 
\[
    p(t) = a \left\{ 1 - r \sin \left(\frac{2\pi}{T_m }(t + \theta_0) \right) \right\} \mathrm{e}^{-(t-t_0)/\tau},
\]
where the time of the original post is assumed to be $t_0= 0$ and $T_{m}= 24$ hours is the period of oscillation. 
The parameters, $a, r, \theta_0$, and $\tau$, correspond to the intensity, the relative amplitude, the phase of the oscillation, and the time constant of decay, respectively.
The memory kernel $\phi(t)$ represents the probability distribution for the reaction time of a follower. 
A heavy-tailed distribution was adopted for the memory kernel~\cite{tideh,seismic}
\[
	\phi (s)= \left\{ \begin{array}{ll}
	c_{0} & (0 \leqq s \leqq s_{0}) \\
	c_{0}(s/s_{0})^{-(1 + \gamma)}  & ({\rm Otherwise}) \\
	\end{array} \right.
\]
The parameters were set to $c_0= 6.94 \times 10^{-4}$ (/seconds), $s_{0} = 300$ seconds, and $\gamma = 0.242$.

\subsection{Proposed model of the spread of fake news} 
We formulate a point process model for the spread of a fake new item. 
Let us assumes that the spread consists 
of two cascades, namely, the one owing to the original news item and the other owing to the correction of the news item. 
The activity of the fake news cascade can be written as the sum of two cascades using TiDeH
\begin{equation}
	\lambda_{\rm prop} (t) = p_1(t) h_1(t) +  p_2(t) h_2(t). 
	\label{eq:TiDeH_Correct}
\end{equation}
The first term $p_1(t) h_1(t)$ represents the rate of the cascade caused by the original news item. 
\begin{eqnarray}
    p_1(t) = 	a_1 \left\{1 + r \sin \left(\frac{2\pi}{T_m }(t + \theta_0) \right)  \right\} \mathrm{e}^{-t/\tau_1},  \ 
	h_1(t) = 	\sum_{i:t_i < \min (t, t_c)} d_i \phi (t-t_{i}), \ 
	\label{eq:1st_Stage}
\end{eqnarray}
where $a_1$ represents the impact of the original news item on the spreading, $\tau_1$ is the decay time constant, $\min(t, t_c)$ represents the smaller of the two values ($t$ or $t_c$), and $t_c$ is the correction time of the fake news item. 
The second term $p_2(t) h_2(t)$ represents the cascade induced by the correction.
\begin{eqnarray}
	p_2(t) = 	a_2 \left \{1 + r \sin \left( \frac{2\pi}{T_m }(t + \theta_0) \right)  \right\}\mathrm{e}^{-(t-t_c)/\tau_2}, \ 
	h_2(t) =    \sum_{i: t_c< t_i <t } d_{i} \phi (t-t_{i}), \  
	\label{eq:2nd_Stage}
\end{eqnarray}
where $a_2$ represents the impact of the falsity of the news on the spreading, and $\tau_2$ is the decay time constant. It is assumed that the circadian parameters of $p_2(t)$ are the same as those of $p_1(t)$. 
Mathematically, the proposed model includes TiDeH as a special case. 
Let us consider the proposed model that satisfies the following conditions 
\begin{eqnarray}
	\tilde{a}= a_1= a_2 e^{-t_c/\tilde{\tau} }, \quad 
	\tilde{\tau}= \tau_1= \tau_2.   \label{eq:Spec_Cond}
\end{eqnarray}
We can see that the proposed model is equivalent to TiDeH (with parameters $a= \tilde{a}$ and $\tau=  \tilde{\tau}$) by substituting Eq. (\ref{eq:Spec_Cond}) into Eqs (\ref{eq:TiDeH_Correct}), (\ref{eq:1st_Stage}), and (\ref{eq:2nd_Stage}). 

\section{Parameter fitting}  \label{sec:fitting}
Here, we describe the procedure for fitting the parameters from the event time series (e.g., the tweeted times).
Seven parameters $\{ a_1, \tau_1; a_2, \tau_2; r, \theta_0; t_c\}$ were determined by maximizing the log-likelihood function 
\begin{equation}
	l= \sum_i  \log \lambda(t_i) - \int_0^{T_{\rm obs} } \lambda(s) ds,
\end{equation}
where $t_i$ is the $i$-th tweeted time, $\lambda(t)$ is the intensity given by Eq. (\ref{eq:TiDeH_Correct}), and $T_{\rm obs}$ is the observation time. 
We first fix the correction time $t_c$ and the other parameters are optimized using the Newton method~\cite{newton}, provided by Scipy~\cite{scipy}, within a range of $12 < \tau_1, \tau_2 <2 T_{\rm obs}$ (hours). 
The correction time is separately optimized using Brent's method \cite{Brent1973} within a range of $0.1 T_{\rm obs} < t_c <0.9T_{\rm obs}$. 
The code for fitting parameters from the tweeted times is available in Github \cite{github_TM}. 

\vspace{1cm}
\begin{figure}[th]
\includegraphics[width=12cm]{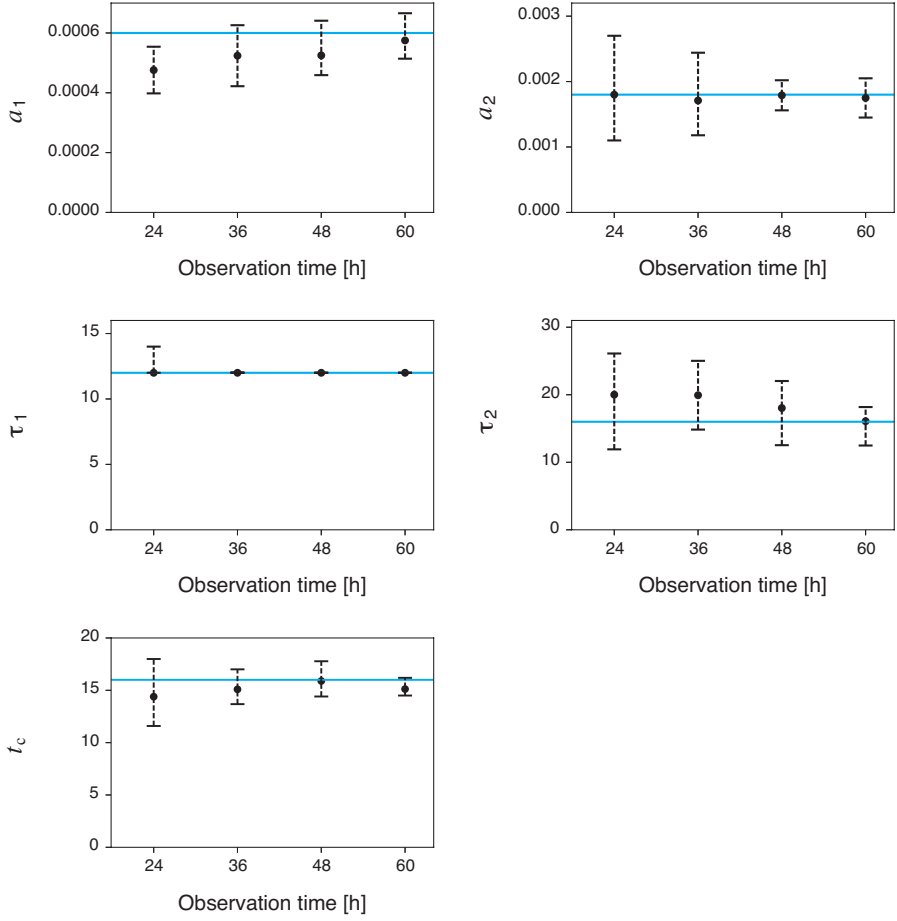}
\caption{
Dependence of the estimation accuracy of parameters $\{ a_1, \tau_1; a_2, \tau_2; t_c \}$ on the observation time. 
Black circles and error bars represent the median and interquartile ranges of the estimates obtained from 100 synthetic data. 
Cyan lines indicate the true value: 
$a_1= 0.0006$, $a_2 = 0.0018$, 
$\tau_1= 12$, $\tau_2= 16$, and $t_c= 16$. 
}
\label{sim_obs}
\end{figure}
\vspace{1cm}

\begin{figure}[ht]
\includegraphics[width=12cm]{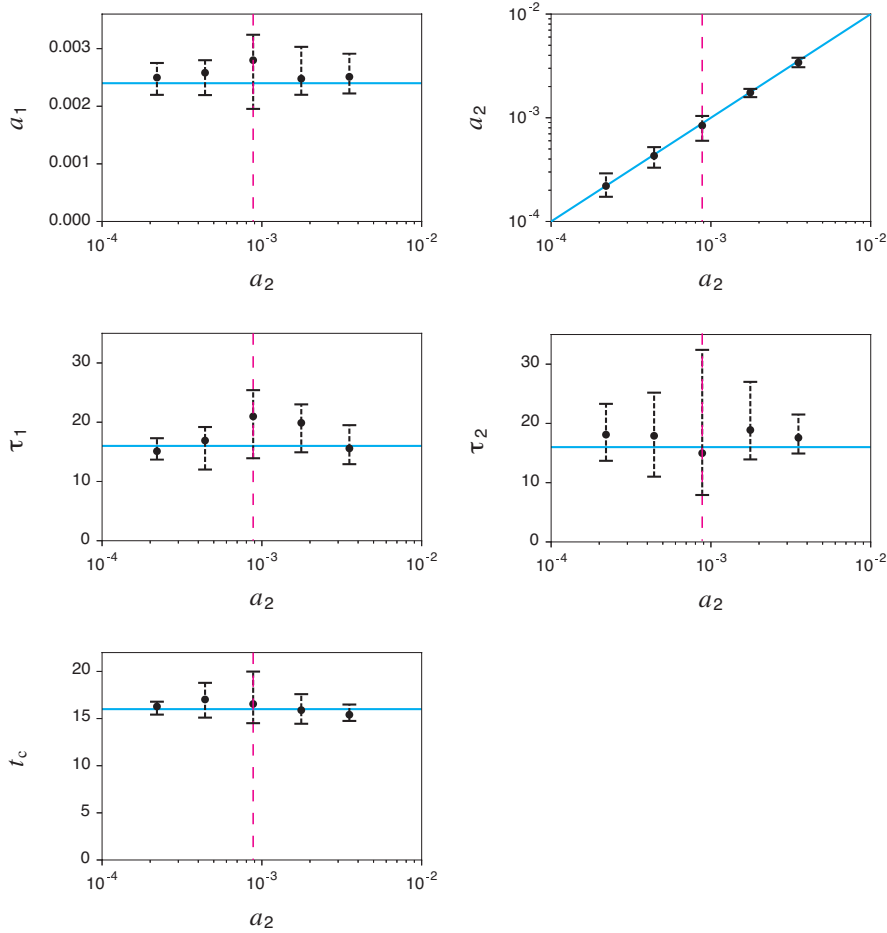}
\caption{
Estimation accuracy of parameters around the non-identifiable domain. 
Black circles and error bars represent the median and interquartile ranges of the estimates obtained from 100 synthetic data. 
Dashed magenta lines represent the non-identifiable domain satisfying $a_2= a_1 e^{-t_c/\tau_2}$. 
Cyan lines indicate the true value: 
$a_1= 0.0024$, $\tau_1= \tau_2= 16$, and $t_c= 16$, and $a_2$ is changed between $2.2 \times 10^{-4}$ and $3.5 \times 10^{-3}$.
}
\label{obs_non-ident}
\end{figure}

We validate the fitting procedure by applying synthetic data generated by the proposed model (Eq. \ref{eq:TiDeH_Correct}).
Figure~\ref{sim_obs} shows the dependence of the estimation accuracy on the observation time $T_{\rm obs}$. 
To evaluate the accuracy, we calculated the median and interquartile ranges of the estimates from 100 trials. 
The estimation error decreases as the observation time increases. 
The result suggests that this fitting procedure can reliably estimate the parameters for sufficiently long observations ($\geq 36$ hours). The medians of the absolute relative errors obtained from 36 hours of synthetic data are 18\%, 11\%, 38\%, 38\%, and 10\% for $a_1$, $\tau_1$, $a_2$, $\tau_2$, and $t_c$, respectively. 
The estimation accuracy of the second cascade parameters ($a_2, \tau_2$) is worse than that of the first cascade parameters ($a_1, \tau_1$). This seems to be caused by the insufficiency of the observed data. While the first cascade parameters are estimated from the entire data, the second cascade parameters are estimated from the observation data after the correction time $t_c$. 
Moreover, the model parameters are not identifiable~\cite{Raue2009,Gontier2020} in the case of $a_1= a_2 e^{-t_c/\tau_2}$ and $\tau_1= \tau_2$. 
Because the proposed model is equivalent to TiDeH ($a_2= 0$, $t_c \geq T_{\rm obs}$) in this case, other parameter sets can also reproduce the observed data.  
Figure~\ref{obs_non-ident} shows that the fitting procedure can estimate the parameters accurately except for the non-identifiable domain.

\vspace{1cm}

\section{Dataset}
We evaluate the proposed model and examine the correction time of fake news based on two datasets of the spread of fake news items. 
Datasets of the spread of fake news based on retweets 
of the original news post~\cite{fakenewsnet,rnn_fake} are publicly available. 
However, rather than a simple retweet, the information sharing of fake news can be complex. 
To cover the information spread in detail, we manually compiled two datasets of fake news items 
spread on Twitter. 
In our dataset, 61\% and 20\% of the tweets are 
retweets of original posts 
in the Recent Fake News dataset and the 2011 Tohoku Earthquake and Tsunami dataset, respectively. 
\vspace{1cm}

\begin{table}[h]
\begin{minipage}{\textwidth}
\center
\scriptsize
\caption{Recent Fake News (RFN): Details of 6 U.S. fake news items}
\scalebox{0.8}{
\begin{tabular}{l|l|r|r|r} \hline
    News No. & Title & Date & No. Posts & $T_{\rm max}$\\ \hline
    \multirow{2}{*}{a. Abolish} & America came along as the first country & \multirow{2}{*}{2019-03-21} & \multirow{2}{*}{1159} & \multirow{2}{*}{36}\\
     & to end (slavery) within 150 years.\footnote{Verbatim quote from Katie Pavlich on Politifact.com, March 19, 2019.}  & & &\\
     \multirow{2}{*}{b. Notredame} & A video clip from the Notre Dame cathedral fire shows a man & \multirow{2}{*}{2019-04-16} & \multirow{2}{*}{1641} & \multirow{2}{*}{132}\\
     & walking alone in a tower of the church ``dressed in Muslim garb." & & &\\
    \multirow{2}{*}{c. Islamic} & Did Ilhan Omar hold `Secret Fundraisers’ & \multirow{2}{*}{2019-03-27} & \multirow{2}{*}{10811} & \multirow{2}{*}{130}\\
     & with `Islamic Groups Tied to Terror’? & & &\\
     \multirow{2}{*}{d. Lionhunter} & Was a trophy hunter eaten alive by lions & \multirow{2}{*}{2019-03-25} & \multirow{2}{*}{25071} & \multirow{2}{*}{88}\\
     & after he killed 3 baboon families? & & &\\
     \multirow{2}{*}{e. Newzealand} & Did New Zealand take Fox News or Sky News off the air & \multirow{2}{*}{2019-03-25} & \multirow{2}{*}{11711} & \multirow{2}{*}{88}\\
     & in response to mosque shooting coverage? & & &\\
     \multirow{2}{*}{f. Sonictrans} & Will the animated character of Sonic the Hedgehog & \multirow{2}{*}{2019-05-06} & \multirow{2}{*}{2319} & \multirow{2}{*}{132}\\
     & be transgender in a new film? & & &\\
     \hline
    \end{tabular}}
    \label{fakenews_data_rnf}
\end{minipage}    
\end{table}

\begin{table}[th]
\center
\scriptsize
\caption{2011 Tohoku earthquake and tsunami (Tohoku): Details of 19 Japanese fake news items}
\scalebox{0.8}{
\begin{tabular}{l|l|r|r|r} \hline
     News No. & Title & Date & No. Posts & $T_{\rm max}$\\ \hline
     \multirow{2}{*}{a. Saveenergy} & \multirow{2}{*}{Large-scale power saving required even in the Kansai region.} & \multirow{2}{*}{2011-03-12} & \multirow{2}{*}{2846} & \multirow{2}{*}{174}\\
     & & & &\\
     \multirow{2}{*}{b. EscapeTokyo} & The bureaucracy in the Ministry of Defense says  & \multirow{2}{*}{2011-03-18} & \multirow{2}{*}{1056} & \multirow{2}{*}{92}\\
     & ``You should escape from Tokyo'' & & &\\
     \multirow{2}{*}{c. Isodin} & \multirow{2}{*}{Isodin is effective against radiation.} & \multirow{2}{*}{2011-03-12} & \multirow{2}{*}{2421} & \multirow{2}{*}{118}\\
     & & & &\\
     \multirow{2}{*}{d. Seaweed} & \multirow{2}{*}{Seaweed is effective against radiation.} & \multirow{2}{*}{2011-03-12} & \multirow{2}{*}{1798} & \multirow{2}{*}{118}\\
     & & & &\\
     \multirow{2}{*}{e. Blog} & \multirow{2}{*}{The blog ``I want you to know what a nuclear plant is.''}  & \multirow{2}{*}{2011-03-13} & 
     \multirow{2}{*}{501} & \multirow{2}{*}{170}\\   &  & & &\\
     \multirow{2}{*}{f. Hutaba} & \multirow{2}{*}{Officials in Hutaba hospital left patients behind and fled.} & \multirow{2}{*}{2011-03-17} & \multirow{2}{*}{1525} & \multirow{2}{*}{118}\\
     &  & & &\\
     \multirow{2}{*}{g. Remark1} & Former chief cabinet secretary Sengoku's remark in Tokushima  & \multirow{2}{*}{2011-03-13} & \multirow{2}{*}{638} & \multirow{2}{*}{170}\\
     &  was inappropriate. & & &\\
     \multirow{2}{*}{h. Remark2} & Former prime minister Hatoyama remarked ``We cannot live & \multirow{2}{*}{2011-03-16} & \multirow{2}{*}{955} & \multirow{2}{*}{120}\\
     & within a 200-kilometer radius of the nuclear power plant.'' & & &\\
     \multirow{2}{*}{i. Visit} & Chief Cabinet Secretary Edano visits Korea a few days  & \multirow{2}{*}{2011-03-15} & \multirow{2}{*}{1973} & \multirow{2}{*}{168}\\
     & after the earthquake.  & & &\\
     \multirow{2}{*}{j. Regulation} & \multirow{2}{*}{Ms. Renho proposes to regulate convenience stores to save energy.}  & \multirow{2}{*}{2011-03-12} & \multirow{2}{*}{7561} & \multirow{2}{*}{156}\\
     & & & &\\
     \multirow{2}{*}{k. Rescue} & \multirow{2}{*}{Ms. Tsujimoto protests U.S. military’s rescue activities.}  & \multirow{2}{*}{2011-03-16} & \multirow{2}{*}{1887} & \multirow{2}{*}{144}\\
     &  & & &\\
     \multirow{2}{*}{l. Taiwan} & \multirow{2}{*}{Taiwan's aid is rejected by the Japanese government.}  & \multirow{2}{*}{2011-03-12} & \multirow{2}{*}{2736} & \multirow{2}{*}{156}\\
     &  & & &\\
     \multirow{2}{*}{m. School seismic} & \multirow{2}{*}{Budget for school seismic retrofitting  was cut by the project screening.}  & \multirow{2}{*}{2011-03-12} & \multirow{2}{*}{1044} & \multirow{2}{*}{174}\\
     & & & &\\
     \multirow{2}{*}{n. Debt} & South Korea asks Japan to borrow money. 
     & \multirow{2}{*}{2011-03-16} & \multirow{2}{*}{399} & \multirow{2}{*}{174}\\
     & Moreover, Japan agrees to this. 
     & & &\\
     \multirow{2}{*}{o. Sanjyo} & Sanjo Junior High School stopped functioning  & \multirow{2}{*}{2011-03-17} & \multirow{2}{*}{379} & \multirow{2}{*}{162}\\
     & due to international students.  & & &\\
     \multirow{2}{*}{p. Fujitv} & \multirow{2}{*}{Japanese TV company Fuji donated to UNICEF Japan.}  & \multirow{2}{*}{2011-03-16} & \multirow{2}{*}{885} & \multirow{2}{*}{124}\\
     &  & & &\\
     \multirow{2}{*}{q. Cartoonist} & \multirow{2}{*}{Japanese cartoonist Mr.Oda donated 1.5 billion yen.}  & \multirow{2}{*}{2011-03-12} & \multirow{2}{*}{2546} & \multirow{2}{*}{171}\\
     & & & &\\
     \multirow{2}{*}{r. Starvation} & \multirow{2}{*}{An infant in Ibaraki died of starvation.}  & \multirow{2}{*}{2011-03-16} & \multirow{2}{*}{2025} & \multirow{2}{*}{144}\\
     & & & &\\
     \multirow{2}{*}{s. Turkey} &  \multirow{2}{*}{Turkey donates 10 billion yen for Japan.}  & \multirow{2}{*}{2011-03-12} & \multirow{2}{*}{2380} & \multirow{2}{*}{158}\\
     & & & &\\ \hline
    \end{tabular}}
    \label{fakenews_data_tohoku}
\end{table}
\vspace{1cm}

\subsection{Recent Fake News (RFN)}
We collected the spread of 10 fake news items from two fact-checking sites, Politifact.com~\cite{politifact} and Snopes.com~\cite{snopes} between March and May, in 2019. 
PolitiFact is an independent, non-partisan site for online fact-checking, mainly for U.S. political news and politicians' statements. 
Snopes.com, one of the first online fact-checking websites, handles political and other social and topical issues. 
Using the Twitter API, tweets highly relevant to the fake news stories were crawled based on the keywords and the URLs. 
We selected six fake news stories based on two conditions: 
1) the number of posts must be greater than 300 and 
2) the observation period must be longer than 36 hours 
(as indicated by the experiments conducted on synthetic data, Fig.~\ref{sim_obs}).  
A summary of the collected fake news stories is presented in Table~\ref{fakenews_data_rnf}.

\subsection{Fake news on the 2011 Tohoku earthquake and tsunami (Tohoku)}
Numerous fake news stories emerged after the 2011 earthquake off the Pacific coast of Tohoku~\cite{tohoku1,hashimoto2020}. 
We collected tweets posted in Japanese from March 12 to March 24, 2011, by using sample streams from the Twitter API. 
There were a total of 17,079,963 tweets. 
We first identified 80 fake news items based on a fake news verification article~\cite{blogos} and obtained the keywords and related URLs of the news items. 
Then, we extracted the tweets highly relevant to the fake news. 
Finally, we selected 19 fake news stories using the same conditions as in the RFN dataset. 
A summary of the collected fake news items is presented in Table~\ref{fakenews_data_tohoku}.

\section{Experimental evaluation}
To evaluate the proposed model, we consider the following prediction task: 
For the spread of a fake news item, we observe a tweet sequence $\{t_i, d_i \}$ up to time $T_{\rm obs}$ from the original post ($t_0= 0$), where $t_{i}$ is the $i$-th tweeted time, $d_{i}$ is the number of followers of the $i$-th tweeting person, and $T_{\rm obs}$ represents the duration of the observation. 
Then, we seek to predict the time series of the cumulative number of posts related to the fake news item during the test period $[T_{\rm obs}, T_{\rm max}]$, where $T_{\rm max}$ is the end of the period. 
In this section, we describe the experimental setup and the proposed prediction procedure, and compare the performance of the proposed method with state-of-the-art approaches.

\subsection{Setup}
The total time interval $[ 0, T_{\rm max}]$ was divided into the training and test periods. 
The training period was set to the first half of the total period $[0, 0.5 T_{\rm max}]$ and the test period was the remaining period $[0.5 T_{\rm max}, T_{\rm max}]$. 
The prediction performance was evaluated by the mean and median absolute error between the actual time series and its predictions: 
\[ {\rm Mean\ Absolute\ Error}= \frac{1}{n_b} \sum_{k=1}^{n_b} |\hat{N}_k- N_k|, \] 
\[ {\rm Median\ Absolute\ Error}= {\rm Median} (|\hat{N}_k- N_k|) \quad (k=1,2,\cdots n_b),  \] 
where $\hat{N}_k$ and $N_k$ are the predicted and actual cumulative numbers of tweets in a $k$-th bin $[ (k-1) \Delta +T_{\rm obs}, k \Delta +T_{\rm obs}]$, respectively, $n_b$ is the number of bins, and $\Delta= 1$ hour is the bin width.

\subsection{Prediction procedure based on the proposed model}
First, we fit the model parameters using the maximum likelihood method from the observation data (see Section 4). 
Second, we calculate the intensity function $\hat{\lambda}(t)$ during the prediction period $t \in [T_{\rm obs}, T_{\rm max}]$
\begin{equation}
    \hat{\lambda}_{\rm prop}(t) = \hat{\lambda}_1(t) + \hat{\lambda}_2(t)
\end{equation}
with 
\begin{equation}
    \hat{\lambda}_1(t)= p_1(t) \sum_{i: t_i<t_c} d_i \phi(t-t_i),
\end{equation}
where $\hat{\lambda}_1(t)$ and $\hat{\lambda}_2(t)$ are the intensities of the first and second cascades, respectively. 
The intensity due to the original news item $\hat{\lambda}_1(t)$ is calculated using the fitted parameters $\{ a_1, \tau_1; r, \theta_0 \}$ and the observations $\{ t_i, d_i \}$ before the inferred correction time $t_c$. 
The number of followers was fixed as 1 ($d_i= 1$) for the Tohoku dataset, because the follower information was not available in the data. 
The intensity due to the correction $\hat{\lambda}_2(t)$ is given by the solution of the integral equation: 
\begin{equation}
    \hat{\lambda}_2(t)= f(t)+ d_p p_2(t) \int_{T_{\rm obs} }^t \hat{\lambda}_2(s) \phi(t-s) ds,
\end{equation}
where 
\[
    f(t)= p_2(t) \sum_{i: t_c< t_i<T_{\rm obs} } d_i \phi(t-t_i),
\]
and $d_p$ is the average number of followers during the observation period.

\subsection{Prediction results}
We evaluated the prediction performance of the proposed model and compared it with three baseline methods: linear regression (LR)~\cite{predict_online}, reinforced Poisson process (RPP)~\cite{rpp} and TiDeH~\cite{tideh}. 
We used the Python 
code in Github~\cite{github_RK} to implement TiDeH. 
Details of the LR and RPP methods are summarized in 
the Appendix (Supporting information S1). 
Figure~\ref{predict_fig} shows three examples of the time series of the cumulative number of posts related to fake news items and their prediction results. 
The proposed method (Fig.~\ref{predict_fig}: magenta) follows the actual time series more accurately than the baselines. 
While the proposed method reproduces the slowing-down 
effect in the posting activity, the baseline models tend to over-estimate the number of posts.

\vspace{1cm}
\begin{figure}[h!]
  \includegraphics[width=12cm]{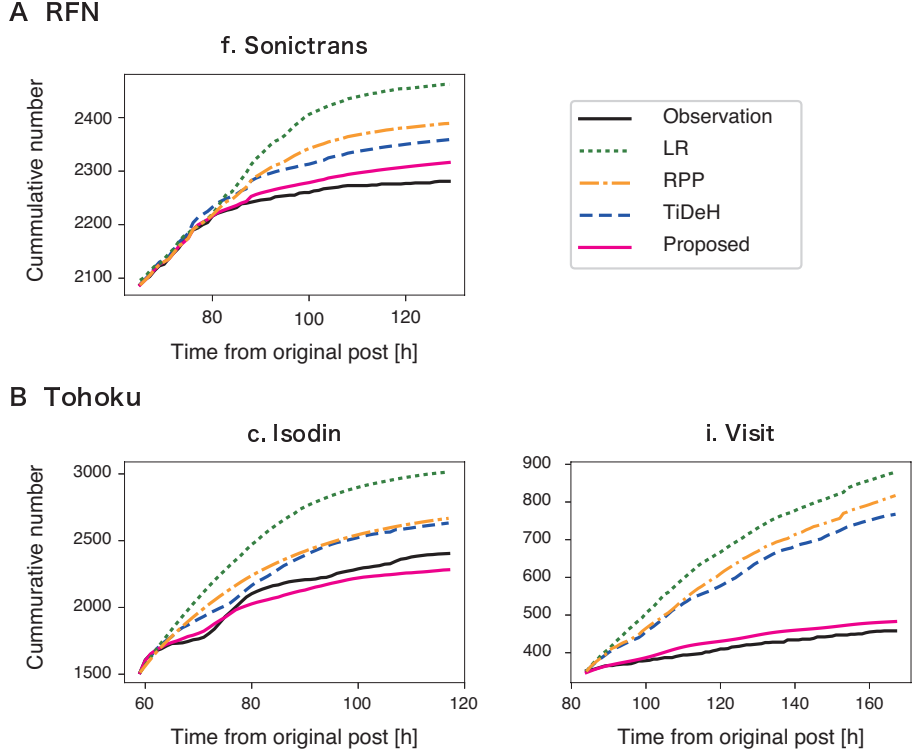}
  \caption{
  Predicting time series of the cumulative number of posts related to a fake news item. 
  Prediction results from (A) RFN and (B) Tohoku datasets are shown. 
  Green, orange, and blue dashed lines represent the prediction results of the baselines (LR, RPP, and TiDeH, respectively). 
  The black and magenta lines represent the observations and their prediction results of the proposed model. 
  }
\label{predict_fig}
\end{figure}
\vspace{1cm}

Next we examine the distribution of the proposed model’s parameters. 
The spreading effect of the falsity of the news item $a_2$ is weaker than that of the news story itself $a_1$ for most fake news items (67\% and 79 \% in the RFN and Tohoku datasets, respectively). 
The result can be attributed to the fact that 
the news story itself is more surprising for the users than the falsity of the news. 
The decay time constant of the first cascade $\tau_1$ is approximately 40 (hours) in both datasets: the median (interquartile range) was 35 (22$-$92) hours and 40 (19$-$54) hours for the RFN and Tohoku datasets, respectively. 
The time constant of the second cascade $\tau_2$ is widely distributed in both datasets, which is consistent with the result observed in the synthetic data (Figure 2). 
The correction time $t_c$ tends to be around 30$-$40 hours after the original post: 32 (21$-$54) hours and 37 (31$-$61) hours for the RFN and Tohoku datasets, respectively. 
A previous study~\cite{Shao2016} reported that the fact-checking sites detect the fake news in 10$-$20 hours after the original post. The result implies that Twitter users recognize the falsity of a fake news item after 10$-$20 hours from the initial report by the fact-checking sites. 

Finally, we evaluated the prediction performance using the two fake news datasets (Table~\ref{experimental_r}). 
Table~\ref{experimental_r} demonstrates that the proposed method outperforms the baseline methods in both datasets and metrics. 
Comparison of the mean error for the proposed model and TiDeH suggests that the two-stage spreading mechanism reduces the mean error by 32 \% and 42 \% in the RFN and Tohoku datasets, respectively. 
Consistent with previous studies~\cite{tideh,seismic}, the methods based on the point process model (the proposed method, TiDeH, and RPP) perform better than the linear regression (LR) method. 
Indeed, the proposed model performs best for most fake news items (100\% and 89\% in the RFN and Tohoku datasets, respectively). 
While TiDeH performs better than the proposed model for the other dataset (8\%), the proposed model still performs much better than the other baselines (RPP and LR). 
Furthermore, we evaluated the goodness-of-fit of the model using Akaike’s information criterion (AIC)~\cite{Akaike}. 
Comparison of AIC values implies that the proposed model achieves a better fit than TiDeH for most fake news items (100\% and 89\% in the RFN and Tohoku datasets, respectively). 
These results suggest that the fake news occasionally spreads in a single cascade rather than in two cascades. 
This might happen when the users already know the falsity of the news in advance (e.g., April Fool's Day) or they are not interested in the falsity of the news at all.  
Overall, these results show that the proposed method is effective for predicting the spread of fake news posts on Twitter.

\vspace{0.5cm}
\begin{table}[h!]
\caption{Prediction performance on the two datasets: mean and median absolute errors per hour. The best results are shown in bold for each case.}
\vspace{0.5cm}
    \centering
    \begin{tabular}{l|p{4em}|p{4em}||p{4em}|p{4em}}\hline
         Datasets & \multicolumn{2}{c||}{RFN} & \multicolumn{2}{c}{Tohoku} \\ \hline
         Metric   & Mean  & Median  & Mean   & Median \\ \hline
         LR       & 88.3  & 5.08    & 13.9   & 4.51\\
         RPP      & 61.8  & 3.12    & 8.23   & 2.30\\
         TiDeH    & 54.2  & 1.89    & 4.12   & 1.99\\
         Proposed  & \textbf{36.9} & \textbf{1.37} & \textbf{2.40} & \textbf{1.80} \\ \hline
    \end{tabular}
    \label{experimental_r}
\end{table}

\section{Inferring the correction time}
We have demonstrated that the proposed method outperforms the existing methods for predicting the evolution of the spread of a fake news item. 
The proposed model assumes that Twitter users realize the falsity of the news around the correction time $t_c$. 
In this section, we examine the validity of this assumption through text mining. 

\begin{figure}[h!]
 \includegraphics[width=12cm]{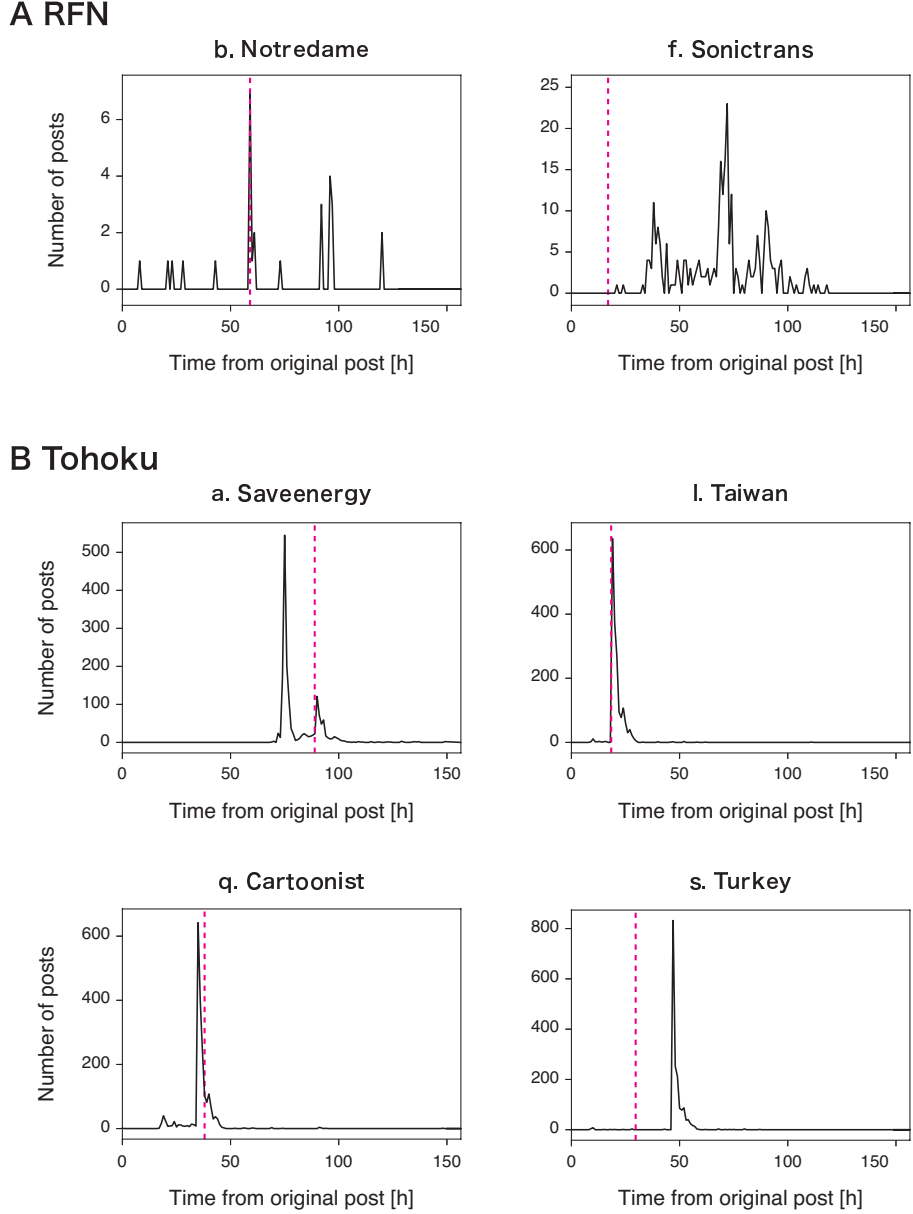}
  \caption{
  Time series of the fake word frequency for fake news items: (A) RFN and (B) Tohoku datasets. 
  In each panel, the black line represents the time series of the ``fake’’ word count per hour for the tweets related to the fake news item and the magenta vertical lines represent the correction time $t_c$.
} 
  \label{Fig:Tc_FakeWords}
\end{figure}

First, we compared the frequency of fake words with inferred correction time $t_c$ (Figure~\ref{Fig:Tc_FakeWords}). 
The fake word frequency is regarded as the number of the tweets having fake words (e.g., false rumors, fake, not true, and not real) in each hour. 
The spread of fake news items in the RFN dataset contained fewer ``fake’’ words than those in the Tohoku dataset: 
29 and 277 fake words in the tweets of b. Notredome and f. Sonictrans in the RFN dataset, and 
1,752, 1,616, 1,723, and 1,930 fake words in the tweets of a. Saveenergy, l. Taiwan, q. Cartoonist, and s. Turkey in the Tohoku dataset during the observation period (150 hours), respectively. 
This is because most of the tweets in the RFN dataset are retweets of the original post.  
We observed that the fake words were posted around the correction time. The peak of the fake word frequency is close to the correction time for Taiwan and Cartoonist in the Tohoku dataset  (Figure~\ref{Fig:Tc_FakeWords}). 

\begin{figure}[h!]
 \includegraphics[width=12cm]{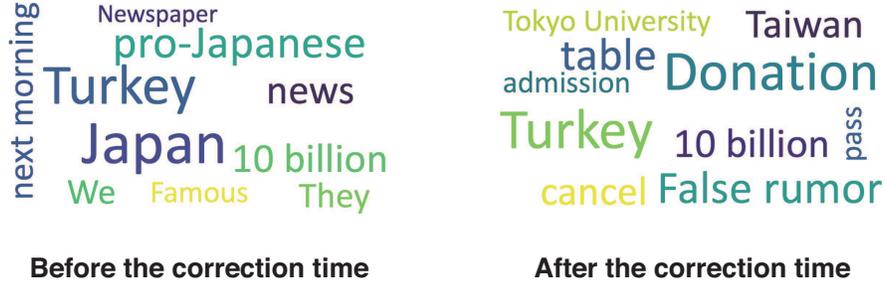}
  \caption{
  Example of word cloud before (left) and after (right) the correction time $t_c$. Each cloud shows the top 10 most frequent words in the fake news story (Turkey in the Tohoku dataset).  
  }
  \label{fig:Tc_WordCould}
\end{figure}

Next, we compared the word cloud before and after the correction time $t_c$. 
Figure~\ref{fig:Tc_WordCould} demonstrates an example of a fake news item spreading ``Turkey'' in the Tohoku dataset. The fake news story is about the huge financial support (10 billion yen) from Turkey to Japan. 
The word cloud before the correction time implies that this fake news item spread due to the fact that Turkey is considered as a pro-Japanese country. 
The term ``False rumor'' starts to appear frequently after the correction time. The word ``Taiwan'' also appears after the correction time, which is related to another fake news story about Taiwan. 
These results suggest that Twitter users realize the falsity of the news after the correction time, which supports the key assumption of the proposed model.

\vspace{1cm}
\vspace{1cm}

\vspace{1cm}

\section{Conclusion}
We have proposed a point process model for predicting the future evolution of the spreading of fake news 
on Twitter (i.e., tweets and re-tweets related to a fake news story). 
The proposed model describes the fake news spread as a two-stage process. First, a fake news item spreads as an ordinary news story. Then, the users recognize the falsity of the news story and spread it as another news story.  
We have validated this model by compiling two datasets of fake news items spread on Twitter. 
We have shown that the proposed model outperforms the state-of-the-art methods for accurately predicting the spread of fake news items. 
Moreover, the proposed model was able to infer the correction time of the news story. Our results based on text mining indicate that Twitter users realize the falsity of the news story around the inferred correction time. 

There are several interesting directions for future works. 
The first direction is to investigate cascades exhibiting multiple bursts. While most fake news cascades exhibit the two-stage spreading pattern, this pattern can also be observed associated with cascades in general. 
A previous study \cite{cascade_recur} found that the cascades of image memes in Facebook consists of multiple popularity bursts and argued that the content virality is the primary driver of cascade recurrence. 
Our work implies that the change in the perception of the content can be another driver. Additional research is needed to determine whether this hypothesis explains the cascade recurrence better than the content virality or not. 
A second direction would be to extend the proposed model. 
While we simply assumed the two-stage process for the spread of a fake news item, this could be extended to describe the spread of fake news in more detail. For example, we can consider multiple types of tweets or a hidden variable to incorporate a soft switch to the second stage from the first one. 
Another direction would be 
to apply the proposed model to 
the practical problems such as fake news detection and mitigation.
We believe that the proposed model provides an important contribution to the modeling of the spread of fake news, 
and it is also beneficial for the extraction of a compact representation of the temporal information related to the spread of a fake news item.

\clearpage
\appendix

\section*{Supporting information}
\section{Baseline methods}
We summarize the baseline methods for predicting the evolution of the spread of a fake news item: linear regression (LR) and reinforced Poisson process (RPP). 

\subsection*{Linear regression (LR)} 
Linear regression is applied to the logarithm of the cumulative number of posts up to time $t$: 
\[
    \log R_t = \alpha_{t} + \log R(T_{\rm obs}) + \sigma_{t}\xi_{t},
\]
where $R_t$ is the cumulative number of posts at the prediction time $t$, $R(T_{\rm obs})$ is the cumulative number of posts at the observation time $T_{\rm obs}$, and $\xi_{t}$ represents the Gaussian random variable with zero mean and unit variance. 
The parameters $\{ \alpha_t, \sigma^2_t \}$ are estimated by the maximum likelihood method from the training data where the tweet sequence in the entire period is available. 
The cumulative number of posts is predicted by the unbiased estimator
\[
    \hat{R}_t = R(T_{\rm obs})\exp(\hat{\alpha}_t + \hat{\sigma}_t^{2}/2), 
\]
where $\hat{R}_t$ is the prediction of the cumulative number, and $\hat{\alpha}_{t}$ and $\hat{\sigma}_t^{2}$ are the fitted parameters.

\subsection*{Reinforced Poisson process (RPP)} 
RPP is a point process model, similar to TiDeH, where the instantaneous function is written as
\[
    \lambda(t) = cf_{\gamma}(t)r_{\alpha}(R(t)),
\]
where $f_{\gamma}(t) = t^{-\gamma}$ describes the aging effect, and  $r_{\alpha}(R) = \epsilon + \frac{1-e^{-\alpha(R+1)}}{1 - e^{-\alpha}}$ is a reinforcement mechanism associated with the multiplicative nature of the spreading.
The model parameters $\{c, \gamma, \alpha \}$ are determined by the maximum likelihood method. 
The cumulative number of posts is evaluated by the expectation of the RPP model, described as follows: 
\[
\frac{dR}{dt} = \lambda(t)
\]
which can be solved analytically
\[
R(t) = (\log(1 + e^x) - x -\log\tilde{\epsilon} - \alpha)/\alpha,
\]
with 
\[
x(t) = \frac{\tilde{\epsilon}c\alpha (T_{\rm obs}^{1-\gamma} - t^{1-\gamma}) }{(1 - \gamma)(1 - e^{-\alpha})} - (R(T_{\rm obs})+1)\alpha - \log(\tilde{\epsilon} - e^{-\alpha(R(T_{\rm obs}) + 1)}),
\]
and $\tilde{\epsilon} = 1 + \epsilon(1 - e^{-\alpha})$. 
This expression is used to predict the cumulative number.

\begin{acknowledgments}
We thank Takeaki Uno for stimulating discussions and JST ACT-I for providing us the opportunity for this collaboration. 
\end{acknowledgments}

\clearpage
\bibliography{article}

\end{document}